# Understanding Fractal Dimension of Urban Form through Spatial Entropy

Yanguang Chen[1]; Jiejing Wang[2]; Jian Feng[1]

(1. Department of Geography, College of Urban and Environmental Sciences, Peking University, Beijing 100871, P.R. China. 2. Department of Urban Planning and Management, School of Public Administration and Policy, Renmin University of China, Beijing, P.R. China. E-mail: chenyg@pku.edu.cn; wangjiejing1120@foxmail.com; fengjian@pku.edu.cn)

**Abstract**: Spatial patterns and processes of cities can be described with various entropy functions. However, spatial entropy always depends on the scale of measurement, and it is difficult to find a characteristic value for it. In contrast, fractal parameters can be employed to characterize scale-free phenomena. This paper is devoted to exploring the similarities and differences between spatial entropy and fractal dimension in urban description. Drawing an analogy between cities and growing fractals, we illustrate the definitions of fractal dimension based on different entropy concepts. Three representative fractal dimensions in the multifractal dimension set are utilized to make empirical analyses of urban form of two cities. The results show that the entropy values are not determinate, but the fractal dimension value is certain; if the linear size of boxes is small enough (e.g., $<1/2^5$), the linear correlation between entropy and fractal dimension is clear. Further empirical analysis indicates that fractal dimension is close to the characteristic values of spatial entropy. This suggests that the physical meaning of fractal dimension can be interpreted by the ideas from entropy and scales and the conclusion is revealing for future spatial analysis of cities.

# 1. Introduction

"*No one is considered scientifically literate today who does not know what a Gaussian distribution is or the meaning and scope of the concept of entropy. It is possible to believe that no one will be considered scientifically literate tomorrow who is not equally familiar with fractals.*"

--John A. Wheeler (1983)

Entropy has been playing an important role for a long time in both spatial measurements and mathematical modeling of urban studies. When mathematical methods were introduced into geography from 1950s to 1970s, the ideas from system theory were also introduced into geographical research. The mathematical methods lead to computational geography and further geo-computation (GC) science, and the system methods result to geographical informatics and then geographical information science (GISc). Along with system thinking, the concepts of entropy entered geographical analysis (Gould, 1970), and the notion of spatial entropy came into being (Batty, 1974). On the one hand, entropy as a measurement can be used to make spatial analysis for urban and regional systems (Batty, 1974; Batty, 1976; Chen, 2015a; Wilson, 2000); on the other, the entropy maximizing method (EMM) can be employed to constitute postulates and make models for human geography (Anastassiadis, 1986; Bussiere and Snickers, 1970; Chen, 2008; Chen, 2012a; Chen and Liu, 2002; Curry, 1964; Wilson, 1968; Wilson, 1970). Unfortunately, the empirical values of spatial entropy often depend on the scale of measurement, and it is difficult to find certain results in many cases. The uncertainty of spatial entropy seems to be associated with the well-known modifiable areal unit problem (MAUP) (Cressie, 1996; Kwan, 2012; Openshaw, 1983; Unwin, 1996). The essence of geographical uncertainty such as MAUP rests with the scaling invariance in geographical space such as urban form (Chen, 2015b; Jiang and Brandt, 2016).

One of the most efficient approaches to addressing scale-free problems is fractal geometry. Many outstanding issues are now can be resolved due to the advent of the fractal theory (Mandelbrot, 1982). Fractal geometry is a powerful tool in spatial analysis, showing new way of looking urban and regional systems (Batty, 1995; Batty, 2008; Frankhauser, 1998). In a sense, fractal dimension is inherently associated with entropy. On the one hand, the generalized fractal dimension is based on Renyi's entropy (Feder, 1988); on the other, it was demonstrated that Hausdorff's dimension is mathematically equivalent to Shannon's entropy (Ryabko, 1986). If the entropy value is not determinate, it can be replaced by fractal dimension. However, in practice, thing is complicated. When and where we should utilize entropy or fractal dimension to make spatial analysis is pending. Many theoretical and empirical studies should be made before clarifying the inner links and essential differences between entropy and fractal dimension.

Geography is a science on spatial difference, and the reflection of difference in human brain yields information. Information can be measured by entropy and fractal dimension. This paper is devoted to exploring the similarities and differences between entropy and fractal dimension in urban studies. In Section 2, a typical regular growing fractal is taken as an example to reveal the connection and distinction between entropy and fractional dimension. The fractal dimension is actually an entropy-based parameter. In Section 3, two Chinese cities, Beijing and Hangzhou, are taken as examples to make empirical analyses. The linear correlation between entropy and fractal dimension is displayed. However, the entropy values rely heavily on spatial scale of measurement, but fractal dimension values are scale-free parameters. In Section 4, the main points of this work are outlined, the shortcomings of the case analyses are stated, and the more related measurements about entropy and fractal dimension are discussed or proposed. Finally, the paper is concluded by summarizing the principal viewpoints.

## 2. Models

### 2.1 The relation of fractal dimension to entropy

Fractal dimension is a measurement of space-filling extent. For urban growth and form, fractal dimension, including box dimension and radial dimension, can act as two indices. One is the index of uniformity for spatial distribution, and the other is the index of space occupancy indicating land use intensity and built-up extent. What is more, the box dimension is associated with spatial entropy (Chen, 1995), and the radial dimension associated with the coefficient of spatial autocorrelation (Chen, 2013). High fractal dimension suggests low spatial difference and strong spatial correlation. For simplicity, let's see a typical growing fractal, which bear an analogy with urban form and growth (Figure 1). This fractal was proposed by Jullien and Botet (1987) and became well known due to the work of Vicsek (1989), and it is also termed Vicsek's figure or box fractal. Geographers employed it to symbolize urban growth (Batty and Longley, 1994; Chen, 2012b; Frankhauser, 1998; Longley *et al*, 1991; White and Engelen, 1993). Starting from an initiator, a point or a square, we can generate the growing fractal by infinitely cumulating space filling or recursive subdivision of space. It is convenient to compute the spatial entropy and fractal dimension of this kind of fractal objects.

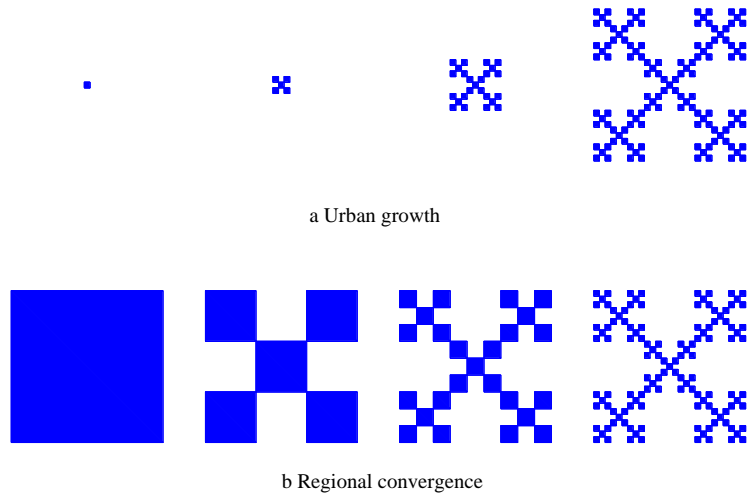

a Urban growth

b Regional convergence

**Figure 1 The growing fractals bearing an analogy to urban growth and regional agglomeration**

(**Note**: The first four steps are displayed. Figure 1(a) reflects a process of infinite space filling, which bears an analogy with urban growth; Figure 1(b) reflects a process of recursive spatial subdivision, which bears an analogy with regional agglomeration such as urbanization.)

The precondition of an index as an effective measurement is that it bears a determinate value. Or else, facing uncertain calculation results, researchers will feel puzzled. However, for the systems without characteristic scale such as fractal cities, we cannot find a determinate entropy value to describe them. In order to compute the spatial entropy, we can use a proper grid to cover a figure. This approach is similar to the box-counting method in fractal studies. A grid consists of a number of squares, which bear an analogy with the boxes for fractal dimension measurement. Sometimes, different types of grid lead to different evaluations. Let's take the well-known box fractal as an example to illustrate the scale dependence of entropy value (Figure 1). If we use a grid comprising a square as a "box" to encompass the fractal object, the "box" just covers the fractal, nothing more, and nothing less. The size of the box is just the measure area of the fractal. Thus, the macro state number and probability are $N=P=1$, the spatial entropy is $H=-P\ln P=\ln N=0$. If the square is averaged into 9 parts, than we have a grid comprising 9 small squares formed by crossed lines. This time, there are 5 nonempty "boxes". The macro state number is $N=5$, the spatial distribution probability is $P=1/N=1/5$, and the spatial entropy is $H=-\sum P\ln P=\ln N=\ln(5)=1.609$ nat. Continuing the 9 small squares into 81 much smaller squares yields 25 nonempty "boxes". Thus the spatial entropy is $H=2*\ln(5)=3.219$ nat, and so on. In short, for the monofractal object, the information entropy equals the corresponding macro state entropy. For the fractal copies with a linear size of $\varepsilon=3^{m-1}$, we have

$$H_e(1/3^{m-1}) = -5^{m-1} * \frac{1}{5^{m-1}} * \ln(\frac{1}{5^{m-1}}) = (m-1)\ln(5) \quad \text{(nat)},$$

in which $m$ represents the step number of fractal generation ($m=1, 2, 3,\ldots$), and $m-1$ denotes the exponent of scale. This suggests that the spatial entropy $H(\varepsilon)$ values depend on the scale of measurement $\varepsilon$. However, if we examine the relationship between the scale series $\varepsilon=1, 1/3, 1/9, \ldots, 1/3^{m-1}$ and nonempty box number series $N(\varepsilon)=1, 5, 25, \ldots, 5^{m-1}$, we will find a power function $N(\varepsilon)=\varepsilon^{-D}$, where the scaling exponent $D=\ln(5)/\ln(3)=1.465$. This exponent value is foreign to the scale $3^{m-1}$. The scaling exponent is just the fractal dimension of the box fractal. The entropy values are indeterminate, but the fractal dimension value is one and only (Table 1).

**Table 1 Entropy and fractal dimension of box fractal based on different scales of measurement**

| Step $m$ | Linear size of fractal copies $\varepsilon_m$ | Number of fractal copies $N_m(\varepsilon)$ | Entropy $H$ (nat) | Fractal dimension $D$ |
|---|---|---|---|---|
| 1 | $1/3^0$ | $1/5^0$ | 0.000 | -- |
| 2 | $1/3^1$ | $1/5^1$ | 1.609 | 1.465 |
| 3 | $1/3^2$ | $1/5^2$ | 3.219 | 1.465 |
| 4 | $1/3^3$ | $1/5^3$ | 4.828 | 1.465 |
| 5 | $1/3^4$ | $1/5^4$ | 6.438 | 1.465 |
| 6 | $1/3^5$ | $1/5^5$ | 8.047 | 1.465 |
| 7 | $1/3^6$ | $1/5^6$ | 9.657 | 1.465 |
| 8 | $1/3^7$ | $1/5^7$ | 11.266 | 1.465 |
| 9 | $1/3^8$ | $1/5^8$ | 12.876 | 1.465 |
| 10 | $1/3^9$ | $1/5^9$ | 14.485 | 1.465 |
| … | … | … | … | … |
| $m$ | $1/3^{m-1}$ | $1/5^{m-1}$ | $(m-1)\ln(5)$ | $\ln(5)/\ln(3)$ |

As indicated above, for a simple fractal object, the macro state entropy based on fractal copy number is equal to the information entropy based on growth probability. The fractal dimension can be defined by the ratio of the state entropy to the logarithm of the linear size of fractal copies. Given a linear size of fractal copies $\varepsilon$ and the number of fractal copies $N(\varepsilon)$, Shannon's information entropy is

$$H(\varepsilon) = -\sum_{i=1}^{N(\varepsilon)} P_i(\varepsilon) \ln P_i(\varepsilon), \tag{1}$$

where $N(\varepsilon)$ denotes the number of fractal copies with linear size $\varepsilon$, $P_i(\varepsilon)$ refers to the probability of growth of the $i$th fractal copy. For the simple regular fractals, the growth probabilities of different fractal copies are equal to one another, i.e., $P_i(\varepsilon)=1/N(\varepsilon)$. Therefore, the macro state entropy equals the information entropy, that is

$$S(\varepsilon) = \ln N(\varepsilon) = H(\varepsilon), \tag{2}$$

in which $S$ indicates the state entropy of urban form. The capacity dimension of fractals is defined based on the state entropy such as

$$D_0 = \frac{S(\varepsilon)}{\ln(1/\varepsilon)} = -\frac{\ln N(\varepsilon)}{\ln \varepsilon}, \tag{3}$$

where $D_0$ denotes the capacity dimension. However, for a complex fractal object, the information entropy is less than the macro state entropy. Based on the information entropy, the information dimension is defined by

$$D_1 = \frac{H(\varepsilon)}{\ln(1/\varepsilon)} = -\frac{\sum_{i=1}^{N(\varepsilon)} P_i(\varepsilon) \ln P_i(\varepsilon)}{\ln(1/\varepsilon)}, \tag{4}$$

where $D_1$ refers to the information dimension.

The state entropy and information entropy can be unified formally. Generalizing varied entropy functions, Renyi (1961) proposed a universal formula to define entropy, which can be expressed as

$$M_q = \frac{1}{1-q} \log \sum_{i=1}^{N(\varepsilon)} P_i^q, \tag{5}$$

where $q$ denotes the order of moments. If $q=0$, $M_0=S$ represents macro state entropy; If $q=1$, $M_1=H$ represents information entropy; if $q=2$, $M_2$ denotes correlation entropy. Accordingly, different types of fractal dimension can be integrated into an expression. Based on the Renyi entropy, the generalized correlation dimension can be defined in the following form (Feder, 1988; Grassberger1985; Mandelbrot, 1999)

$$D_q = -\lim_{\varepsilon \to 0} \frac{H_q(\varepsilon)}{\log \varepsilon} = \frac{1}{q-1} \lim_{\varepsilon \to 0} \frac{\log \sum_{i=1}^{N(\varepsilon)} P_i^q}{\log \varepsilon}, \tag{6}$$

where $D_q$ is the generalized dimension of order $q$. If $q=0$, $D_q=D_0$ refers to *capacity dimension*, if $q=1$, $D_q=D_1$ refers to *information dimension*, and if $q=2$, $D_q=D_2$ refers to *correlation dimension* (Grassberger, 1983). In theory, $q \in (-\infty, +\infty)$. Thus we get a multifractal spectrum based on $q$. for the monofractal phenomena, $D_0=D_1=D_2$, but for the multifractal systems, $D_0>D_1>D_2$.

## 2.2 Entropy and fractal dimension indicating geo-spatial development

Fractal theory suggests new way of academic research, especially in geographical analysis. In future science, culture, and education, fractal concept will play an important role and will become as common as entropy and maps (Batty, 1992; Wheeler, 1983). In fact, entropy can be associated with fractal dimension by both mathematical forms and physical meaning. For a given linear scale $\varepsilon$, the fractal dimension is equivalent to the corresponding entropy. The generic conclusion was drawn by Ryabko (1986), who argued that the Shannon's entropy is equivalent in mathematics to the Hausdorff dimension. The spatial entropy is a measurement of spatial uniformity. Consequently, the fractal dimension is also the uniformity measurement of spatial distribution.

Both entropy and fractal dimension can be employed to describe urban form and growth. Fractal dimension changes can reflect spatial concentration and diffusion. As indicate above, we have two typical approaches to constructing the deterministic fractals. One is to use an iteration procedure, and the other is by subsequent divisions of the original square (Vicsek, 1989). The former process bears an analogy with urban growth, while the latter process has an analogy to regional agglomeration (Figure 1). The same goal can be reached by different routes. That is, the final results are the same with each other, and the fractal dimension is $D=\ln(5)/\ln(3)=1.465$. Now, let's examine the processes of fractal development rather than the ultima results. For the fractal process in Figure 1(a), the initiator is a point with dimension $D=0$, corresponding to the information entropy $H=0$, but the final dimension is $D=1.465$, corresponding to the information entropy $H=1.609$. The dimension value and information entropy go up (from 0 to 1.465, 0 to 1.609). For the fractal process in Figure 1(b), the initiator is a square with dimension $D=2$, corresponding to the information entropy $H=2.197$, but the final dimension is $D=1.465$. The dimension value and information entropy go down (from 2 to 1.465, 2.197 to 1.609). Figure 1(a) suggests a process of spatial spread, while Figure 1(b) implies a process of spatial concentration. The aim of this article is to reveal the numerical relation between spatial entropy and fractal dimension, and real urban systems are more complicated than the regular fractal shown above.

# 3. Empirical analysis

## 3.1 Study area, materials, and method

The spatial entropy and fractal parameters can be employed to make spatial analysis of the urban form and growth. One of the examples is Beijing city, the capital of China. The datasets came from the remote sensing images of five years, that is, 1988, 1992, 1999, 2006, and 2009 (Figure 2). The ground resolution of these images is 30 meters (Chen and Wang, 2013). The functional box-counting method can be used to measure the spatial entropy and fractal dimension (Figure 3). This method was originally adopted by Lovejoy *et al* (1987) to analyze radar rain data, and Chen (1995) improved this method in urban studies by replacing the largest box with arbitrary area with the largest box with a measure area of an urban system. In fact, the functional box-counting method can be termed Rectangle Space Subdivision (RSS) method (Chen and Wang, 2013; Feng and Chen, 2010). The geometrical basis of RSS is the recursive subdivision of space and the cascade structure of hierarchies (Batty and Longley, 1994; Goodchild and Mark, 1987). Its mathematical basis is the logical relationship between the exponential laws based on translational symmetry and the power laws based on dilation symmetry (Chen, 2012a).

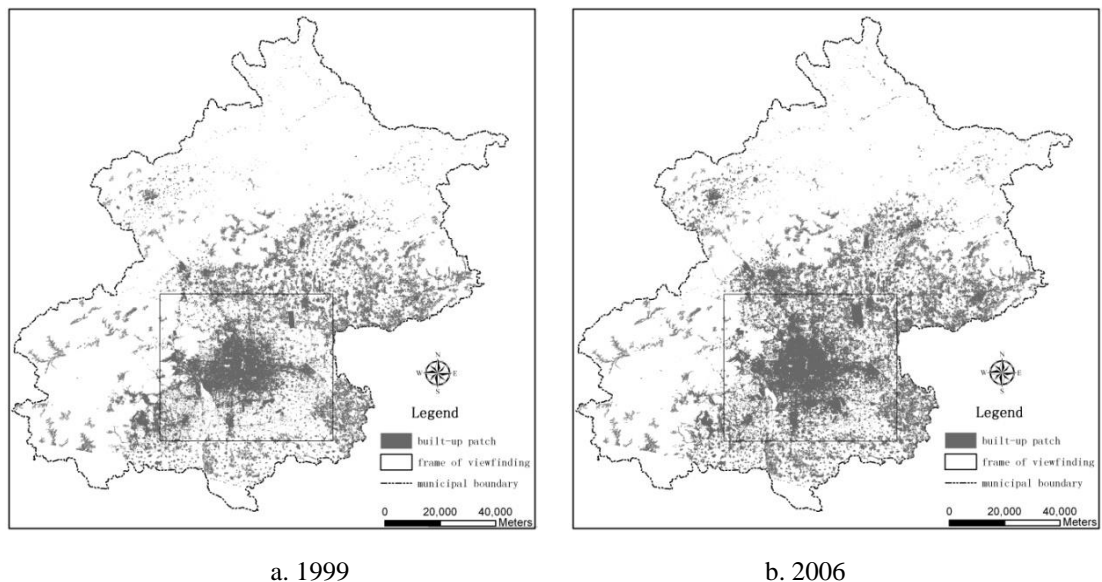

a. 1999　　　　　　　　　　　　　　b. 2006

**Figure 2 Two sketch maps of Beijing's urban form and growth (1992 and 2006 years)**

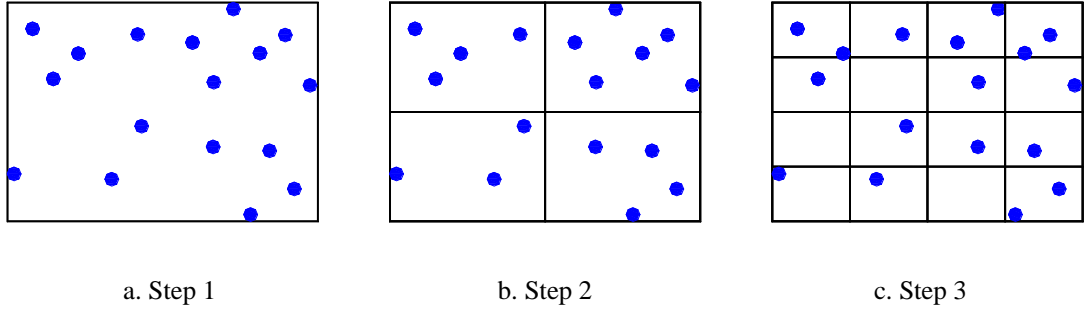

| a. Step 1 | b. Step 2 | c. Step 3 |

**Figure 3 A sketch map of the functional box-counting method for spatial entropy and fractal dimension measurement (first three steps)**

### 3.2 Results and findings

Applying the functional box method above-illustrated to Beijing metropolitan area yields the datasets of the spatial distribution of urban land use in five years. For comparability, the box size is fixed for different years. The area of the largest box equals the measure area of the metropolis in 2009. Using the datasets, we can calculate state entropy, information entropy, and Renyi entropy (Table 2). The relationships between the logarithms of the linear size of box (scale) and the entropy values (measurements) take on a linear trend (Figure 4). The slopes of the trend lines give the capacity dimension $D_0$, information dimension $D_1$, and correlation dimension $D_2$ (Table 3). As indicated above, $D_0$ is based on macro state entropy (Boltzmann entropy), $D_1$ is based on information entropy (Shannon entropy), and $D_2$ is based on generalized entropy (Renyi entropy). The standard errors of all the fractal dimension values are less than 0.04. According to Benguigui *et al* (2000), the fractal structure of Beijing's urban form is very clear.

**Table 2 The state entropy, information entropy, and Renyi entropy of Beijing's urban form**

| Scale ($\varepsilon$) | Three types of entropy values | | | | | | | | |
|---|---|---|---|---|---|---|---|---|---|
| | 1988 | | | 1992 | | | 1999 | | |
| | $q=0$ | $q=1$ | $q=2$ | $q=0$ | $q=1$ | $q=2$ | $q=0$ | $q=1$ | $q=2$ |
| $1/2^0$ | 0.0000 | 0.0000 | 0.0000 | 0.0000 | 0.0000 | 0.0000 | 0.0000 | 0.0000 | 0.0000 |
| $1/2^1$ | 2.0000 | 1.9672 | 1.9364 | 2.0000 | 1.9775 | 1.9569 | 2.0000 | 1.9650 | 1.9279 |
| $1/2^2$ | 4.0000 | 3.5770 | 3.2749 | 4.0000 | 3.4693 | 3.1363 | 4.0000 | 3.6850 | 3.4668 |
| $1/2^3$ | 6.0000 | 5.3986 | 5.0152 | 6.0000 | 5.2657 | 4.8831 | 6.0000 | 5.5608 | 5.2991 |
| $1/2^4$ | 7.9944 | 7.2654 | 6.8736 | 7.9887 | 7.1303 | 6.7571 | 7.9887 | 7.4614 | 7.1875 |
| $1/2^5$ | 9.9233 | 9.1260 | 8.7581 | 9.9352 | 9.0084 | 8.6618 | 9.9556 | 9.3590 | 9.0951 |
| $1/2^6$ | 11.7224 | 10.9502 | 10.6430 | 11.7394 | 10.8623 | 10.5656 | 11.8431 | 11.2301 | 10.9973 |
| $1/2^7$ | 13.3201 | 12.7152 | 12.4926 | 13.3829 | 12.6723 | 12.4488 | 13.5868 | 13.0592 | 12.8761 |
| $1/2^8$ | 14.9124 | 14.4897 | 14.3423 | 14.9744 | 14.4859 | 14.3320 | 15.2798 | 14.8915 | 14.7582 |
| $1/2^9$ | 16.5871 | 16.3073 | 16.2155 | 16.6369 | 16.3234 | 16.2255 | 17.0079 | 16.7416 | 16.6532 |

**Continued Table 2**

| Scale ($r$) | Three types of entropy values | | | | | |
|---|---|---|---|---|---|---|
| | 2006 | | | 2009 | | |
| | $q=0$ | $q=1$ | $q=2$ | $q=0$ | $q=1$ | $q=2$ |
| $1/2^0$ | 0.0000 | 0.0000 | 0.0000 | 0.0000 | 0.0000 | 0.0000 |
| $1/2^1$ | 2.0000 | 1.9964 | 1.9928 | 2.0000 | 1.9942 | 1.9886 |
| $1/2^2$ | 4.0000 | 3.8306 | 3.6976 | 4.0000 | 3.9231 | 3.8617 |
| $1/2^3$ | 6.0000 | 5.7431 | 5.5816 | 6.0000 | 5.8783 | 5.7959 |
| $1/2^4$ | 7.9887 | 7.6778 | 7.5024 | 7.9944 | 7.8342 | 7.7421 |
| $1/2^5$ | 9.9672 | 9.6112 | 9.4351 | 9.9701 | 9.7897 | 9.6896 |
| $1/2^6$ | 11.8986 | 11.5117 | 11.3460 | 11.9263 | 11.7163 | 11.6131 |
| $1/2^7$ | 13.7249 | 13.3678 | 13.2316 | 13.8404 | 13.6112 | 13.5160 |
| $1/2^8$ | 15.5017 | 15.2185 | 15.1195 | 15.7081 | 15.4972 | 15.4202 |
| $1/2^9$ | 17.2877 | 17.0879 | 17.0245 | 17.5632 | 17.3994 | 17.3428 |

**Note**: The base of the logarithm is 2, thus the unit of information quantity is bit. If a calculation is based on natural base of logarithm, all these values should be multiplied by ln(2).

**Table 3 The capacity dimension, information dimension, and correlation dimension of Beijing's urban form and the related statistics**

| Type | Parameter/statistic | 1988 | 1992 | 1999 | 2006 | 2009 |
|---|---|---|---|---|---|---|
| Capacity dimension | Parameter $D_0$ | 1.8507 | 1.8584 | 1.8998 | 1.9297 | 1.9575 |
| | Standard error $\delta$ | 0.0310 | 0.0299 | 0.0222 | 0.0160 | 0.0095 |
| | R Square $R^2$ | 0.9978 | 0.9979 | 0.9989 | 0.9995 | 0.9998 |
| Information dimension | Parameter $D_1$ | 1.8099 | 1.8130 | 1.8602 | 1.8986 | 1.9335 |
| | Standard error $\delta$ | 0.0070 | 0.0114 | 0.0050 | 0.0053 | 0.0051 |
| | R Square $R^2$ | 0.9999 | 0.9997 | 0.9999 | 0.9999 | 0.9999 |
| Correlation dimension | Parameter $D_2$ | 1.8039 | 1.8071 | 1.8530 | 1.8909 | 1.9259 |
| | Standard error $\delta$ | 0.0201 | 0.0271 | 0.0130 | 0.0065 | 0.0027 |
| | R Square $R^2$ | 0.9990 | 0.9982 | 0.9996 | 0.9999 | 1.0000 |

The empirical results support the theoretical inference based on regular fractals. For the urban agglomeration of Beijing, the spatial entropy values depend on the scale of measurements. When the linear size of boxes becomes smaller and smaller, the entropy values become larger and larger. No characteristic entropy value can be found for spatial description. However, there is a determinate relation between the linear sizes of boxes and the corresponding entropy values. By this relation, a number of entropy values can be transformed into a fractal dimension value. In other words, we cannot find a characteristic scale for entropy values, but we can use fractal dimension as a characteristic parameter to reflect spatial structure of a city.

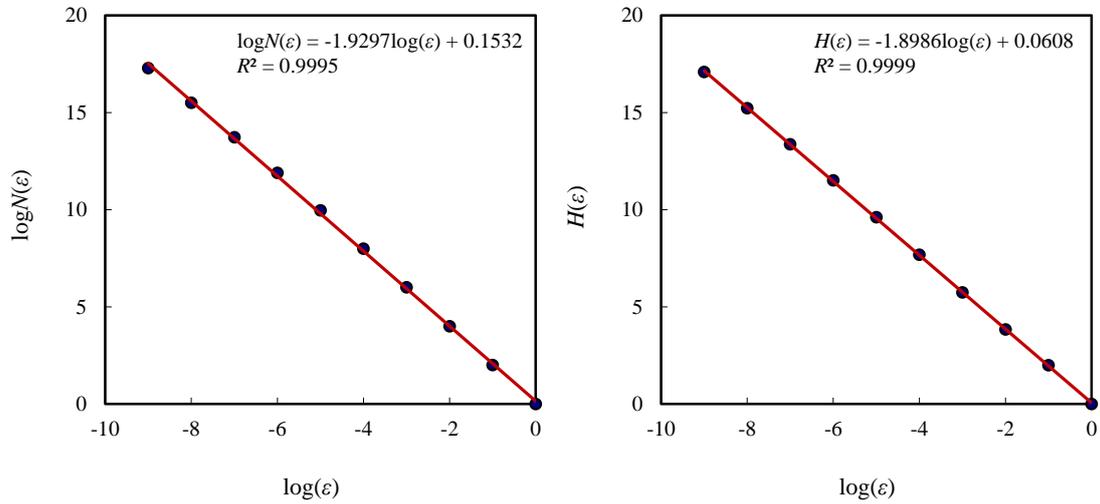

a. Capacity dimension ($q=0$, $D_0$)   b. Information dimension ($q=1$, $D_1$)

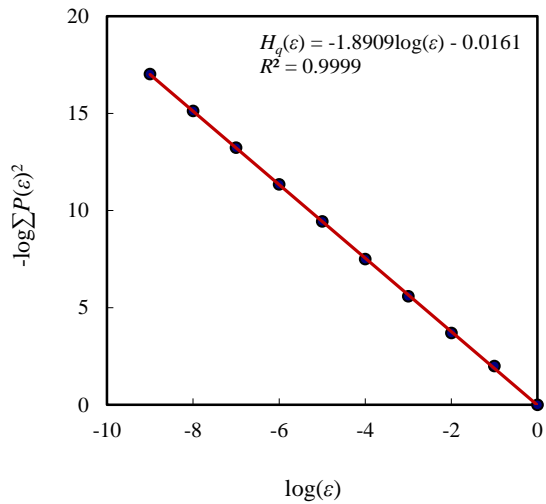

c. Correlation dimension ($q=2$, $D_2$)

**Figure 4 The plots for evaluating multifractal parameters of Beijing's urban form and growth patterns (2006)**

Now, let's examine the correlation relationships between three types of entropy and the corresponding fractal dimensions. As indicated above, for given moment orders ($q$), based on different linear sizes of boxes ($\varepsilon$), we have different entropy values, but the corresponding fractal dimension value does not depend on the linear sizes. Using the datasets of entropy and fractal dimension values, we can calculate the square of correlation coefficients ($R$ squared). The squared $R$ is known as goodness of fit or determination coefficient in linear regression analysis (Table 4). The results show three characters. First, the smaller the linear sizes of boxes, the higher the squared correlation coefficient values; second, the closer to $q=1$ the moment order, the higher the squared correlation coefficient values tend to be; third, there seems to be a limit for the smallest linear size of boxes (Figure 5). The relation between entropy ($M_q$) and fractal dimension ($D_q$) can be expressed as $D_q=aM_q+b$, where $a$ and $b$ are constants. This suggests that the fractal dimension of cities includes the meaning of spatial entropy. If the linear size of spatial measurement is small enough, the entropy and fractal dimension can be replaced with one another in theory, and supplement each other in practice.

Table 4 The squared coefficients of correlation between fractal dimension values and the corresponding entropy values of Beijing's urban form (1988-2009)

| Moment order | Correlation coefficient square ($R^2$) | | | | | | | | |
|---|---|---|---|---|---|---|---|---|---|
| $q$ | $\varepsilon=1/2^1$ | $\varepsilon=1/2^2$ | $\varepsilon=1/2^3$ | $\varepsilon=1/2^4$ | $\varepsilon=1/2^5$ | $\varepsilon=1/2^6$ | $\varepsilon=1/2^7$ | $\varepsilon=1/2^8$ | $\varepsilon=1/2^9$ |
| $q=0$ | -- | -- | -- | 0.0095 | 0.9330 | 0.9786 | 0.9963 | 1.0000 | 0.9990 |
| $q=1$ | 0.5909 | 0.9460 | 0.9549 | 0.9641 | 0.9768 | 0.9877 | 0.9963 | 0.9994 | 0.9999 |
| $q=2$ | 0.5460 | 0.9640 | 0.9780 | 0.9846 | 0.9893 | 0.9925 | 0.9966 | 0.9989 | 0.9996 |

**Note**: If q=0, the first three square correlation coefficient values cannot be calculated.

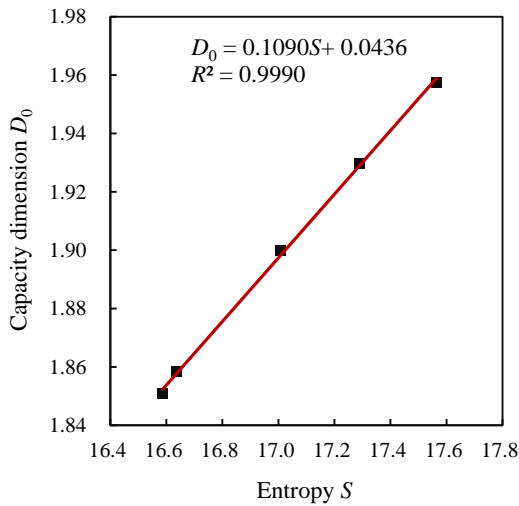 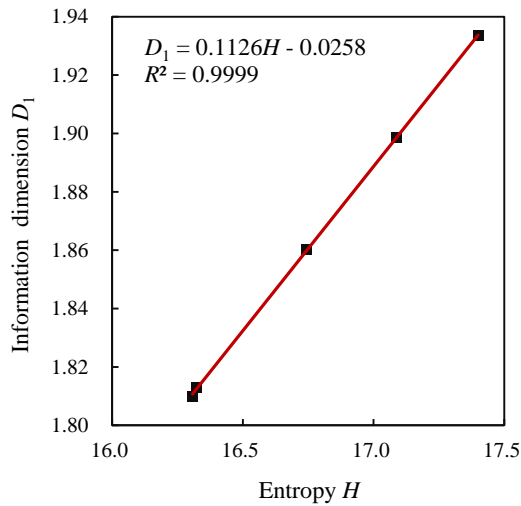

a. Capacity dimension $D_0$ and entropy $H_0$     b. Information dimension $D_1$ and entropy $H_1$

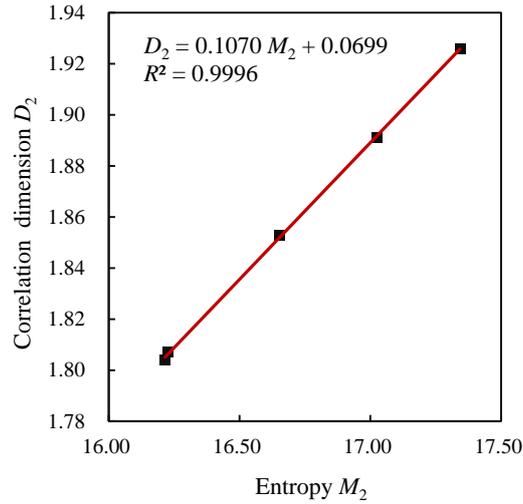

c. Correlation dimension $D_2$ and entropy $M_2$

**Figure 5 The linear relationships between fractal dimension values and the corresponding entropy values (based on the linear size $1/2^9$) of Beijing's urban form (1988-2009)**

### 3.3 Further observational evidences and findings

More empirical evidence can be found to attest the numerical relationships between spatial entropy and fractal dimension. The city of Hangzhou is another typical example. The spatial patterns of Hangzhou's urban land use bear fractal structure, and can be characterized with fractal dimension (Figure 6). Using the functional box-counting method, Feng and Chen (2010) once calculated the capacity dimension of Hangzhou's urban form in four different years (1949, 1959, 1980, and 1996). Different from the case of Beijing city, the variable boxes were employed to make spatial measurement for suiting city sizes in different years.

The analytical process is similar to that is made for Beijing city. Based on the published datasets by Feng and Chen (2010), the macro state entropy can be computed and the coefficient of correlation between state entropy and capacity dimension values can be worked out (Table 5). According to the results, when the linear sizes of boxes become smaller and smaller, the linear relationships between entropy and fractal dimension become clearer and clearer. For the large size of boxes, the regularity does not appear; if the box sizes become small enough, the linear fit of fractal dimension to entropy is close to perfection (Figure 7). The relation between entropy ($S$) and fractal dimension ($D_0$) can be written as $D_0=aS+b$, where *a* and *b* are parameters. This case lends further support the inference that there is potential equivalence of fractal dimension to entropy under certain condition.

Table 5 The macro state entropy, capacity dimension, and the squared coefficient of correlation between the fractal dimension and entropy of Hangzhou's urban form (1949-1996)

| Type | Box size | Parameter and statistic | | | | Squared correlation coefficient $R^2$ |
|---|---|---|---|---|---|---|
| | | 1949 | 1959 | 1980 | 1996 | |
| **Entropy** | $1/2^1$ | 1.3863 | 1.3863 | 1.3863 | 1.3863 | -- |
| | $1/2^2$ | 2.7081 | 2.7081 | 2.7726 | 2.7726 | 0.8992 |
| | $1/2^3$ | 3.6376 | 3.7842 | 4.1431 | 4.1431 | 0.9812 |
| | $1/2^4$ | 4.3820 | 4.7536 | 5.4161 | 5.4765 | 0.9943 |
| | $1/2^5$ | 5.3132 | 5.8201 | 6.6631 | 6.7979 | 0.9981 |
| | $1/2^6$ | 6.3244 | 6.9048 | 7.8793 | 8.0983 | 0.9997 |
| | $1/2^7$ | 7.3896 | 8.0446 | 9.1344 | 9.2927 | 0.9975 |
| | $1/2^8$ | 8.5704 | 9.2595 | 10.3984 | 10.6780 | 0.9999 |
| | $1/2^9$ | 9.8901 | 10.5666 | 11.6846 | 11.9757 | 0.9999 |
| **Fractal dimension** | $D_0$ | 1.5013 | 1.6379 | 1.8581 | 1.9115 | |
| | $\delta$ | 0.9953 | 0.9983 | 0.9995 | 0.9997 | |
| | $R^2$ | 0.0365 | 0.0236 | 0.0148 | 0.0112 | |
| **Eigen entropy** | $b$ | 0.6630 | 0.6095 | 0.5379 | 0.5230 | |
| | $R^2$ | 0.9953 | 0.9983 | 0.9995 | 0.9997 | |
| | $1/b$ | 1.5084 | 1.6406 | 1.8591 | 1.9120 | |

**Note**: In the last column, the values of $R^2$ refer to the squared coefficient of correlation between fractal dimension and entropy based on different linear sizes of boxes; in the last sixth rows, the values of $R^2$ denote the goodness of fit for estimating the fractal dimension. Differing from Feng and Chen (2010), the scaling range is not considered for fractal dimension evaluation in this paper.

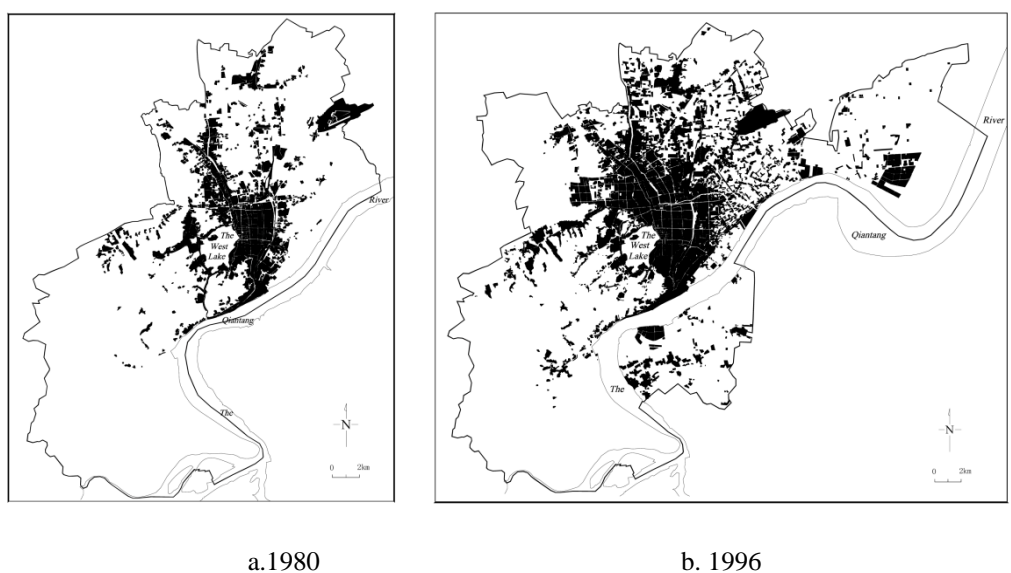

a. 1980   b. 1996

**Figure 6. Two sketch maps of Hangzhou's urban form and growth (1980 and 1996 years)**

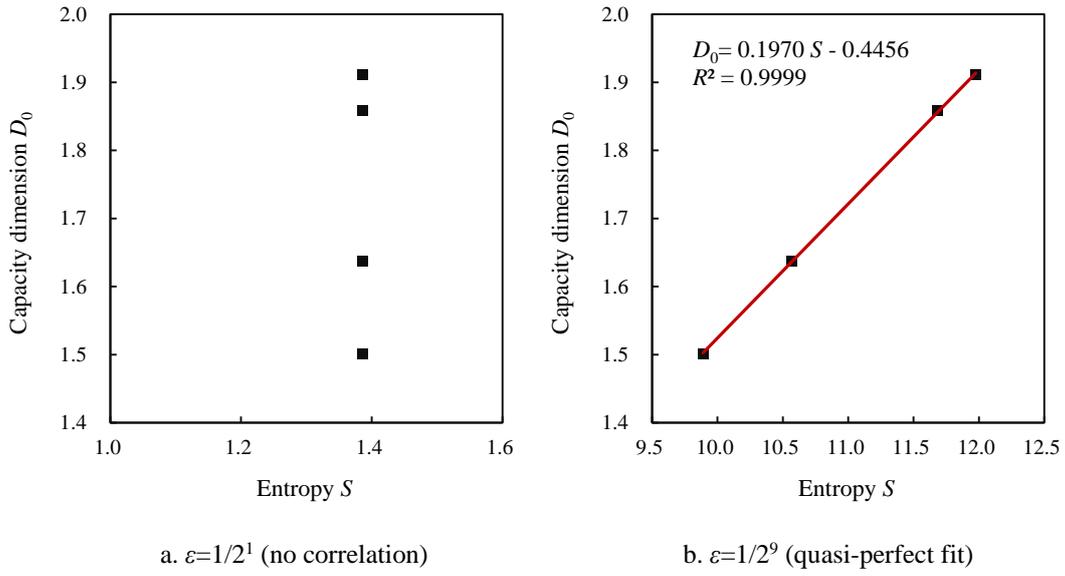

a. $\varepsilon=1/2^1$ (no correlation)    b. $\varepsilon=1/2^9$ (quasi-perfect fit)

**Figure 7 The linear relationships between capacity dimension values and the macro state entropy values (based on the linear size $1/2^9$) of Hangzhou's urban form (1949-1996)**

[**Note**: If the linear size of box is too large, say, 1/2, the correlation cannot appear; if the box size is small enough, say, $1/2^9$, the linear correlation will become very clear.]

However, the urban density of population distribution in Hangzhou can be described with spatial entropy rather than fractal dimension. Urban population density follows Clark's law, which can be expressed a negative exponential function (Clark, 1951). Clark's law can be expressed as below:

$$\rho(r) = \rho_0 e^{-r/r_0}, \tag{7}$$

where $\rho(r)$ denotes the population density at the distance $r$ from the center of city ($r=0$). As for the parameters, $\rho_0$ refers to the urban central density $\rho(0)$, and $r_0$ to the characteristic radius of the population distribution (Batty and Longley, 1994; Takayasu, 1992). The negative exponent distribution differs from the inverse power-law distribution. The former bears characteristic scale $r_0$, while the latter possesses no characteristic scale. For the distribution with characteristic scale, we cannot calculate fractal dimension, but we can compute the spatial entropy using the urban density data. The population density data of Hangzhou city based on four years of census (1964, 1980, 1990, and 2000) was processed by Feng (2002). Using these data sets, we can evaluate the spatial information entropy of urban density distribution (Table 6). The characteristic parameter is the average radius of population distribution ($r_0$), which can be estimated with equation (8). This radius indicates the central density is associated with the maximum entropy (Chen, 2008).

Table 6 The information entropy of urban population density and the related measurements of Hangzhou (1964-2000)

| Year | Average density ($\rho$) | Characteristic radius ($r_0$) | Entropy ($H$) | Entropy ratio ($H/H_{max}$) |
|------|--------------------------|-------------------------------|---------------|------------------------------|
| 1964 | 4721.3643 | 3.5644 | 2.4593 | 0.7548 |
| 1982 | 5702.7587 | 3.6713 | 2.4843 | 0.7625 |
| 1990 | 6774.3742 | 3.6285 | 2.5486 | 0.7822 |
| 2000 | 8411.6099 | 3.9457 | 2.7251 | 0.8364 |

In fact, urban form has no characteristic scale, but spatial entropy bear characteristic values. In other words, despite that spatial entropy values depend on measurement scales, but entropy itself has a characteristic scale. It can be demonstrated that the characteristic value is just the corresponding fractal dimension. This can be understood by analogy with Clark's law above mentioned. In the formulae of fractal dimensions, entropy is a logarithmic function of spatial scales. The inverse function is just exponential function, which bears an analogy with equation (7). In fact, from equation (6) it follows

$$\varepsilon(q) = e^{-H_q/D_q} = ae^{-bH_q}, \qquad (8)$$

in which $a$ refers to proportionality coefficient, and $b=1/D_q$ represents the decay parameter. Equation (9) is identical in form to equation (7). Comparing equation (9) with equation (7) suggests that fractal dimension $D_q$ is just the characteristic value of general entropy $H_q$ relative to the scale $\varepsilon$. This judgment can be verified by the observational data of Beijing city. Taking logarithm of equation (9) gives a linear equation such as $\ln(\varepsilon)=\ln(a)+b*H_q$. Suppose that $H_q$ serves for an independent variable and $\ln(\varepsilon)$ acts as an dependent variable. The least squares calculations based on the linear relation and the data sets in Table 2 yield a series of regression coefficients. The reciprocals of these regression coefficients, $1/b$, indicate the characteristic scales of spatial entropy $H_0$, $H_1$, and $H_2$ (Table 7). The characteristic values of spatial entropy are very close to the fractal dimension values, which are displayed in Tables 3 and 7. The similar method can be applied to the datasets of Hangzhou city, and the results are added to the last three lines of Table 5. The reciprocal values of the parameter b is close to the values of the corresponding fractal dimension $D_0$. These results are important for us to understand and interpret fractal dimension by means of spatial entropy.

Table 7 The characteristic values of spatial entropy (1/*b*) and the corresponding fractal dimension values ($D_q$)

| Year | *a* | *b* | $R^2$ | 1/*b* | $D_q$ |
| --- | --- | --- | --- | --- | --- |
| 1988 | 1.1183 | 0.5391 | 0.9978 | 1.8549 | 1.8507 |
|      | 1.0132 | 0.5525 | 0.9999 | 1.8101 | 1.8099 |
|      | 0.9367 | 0.5538 | 0.9990 | 1.8057 | 1.8039 |
| 1992 | 1.1122 | 0.5370 | 0.9979 | 1.8622 | 1.8584 |
|      | 0.9842 | 0.5514 | 0.9997 | 1.8136 | 1.8130 |
|      | 0.9088 | 0.5524 | 0.9982 | 1.8103 | 1.8071 |
| 1999 | 1.0787 | 0.5258 | 0.9989 | 1.9018 | 1.8998 |
|      | 1.0089 | 0.5375 | 0.9999 | 1.8603 | 1.8602 |
|      | 0.9577 | 0.5395 | 0.9996 | 1.8537 | 1.8530 |
| 2006 | 1.0547 | 0.5179 | 0.9995 | 1.9308 | 1.9297 |
|      | 1.0222 | 0.5267 | 0.9999 | 1.8987 | 1.8986 |
|      | 0.9938 | 0.5288 | 0.9999 | 1.8911 | 1.8909 |
| 2009 | 1.0324 | 0.5108 | 0.9998 | 1.9578 | 1.9575 |
|      | 1.0229 | 0.5172 | 0.9999 | 1.9336 | 1.9335 |
|      | 1.0109 | 0.5192 | 1.0000 | 1.9260 | 1.9259 |

**Note**: Before making the least squares regression, all the entropy values on the base of 2 in Table 2 have been transformed into the entropy values on the natural base, *e*. the characteristic values of spatial entropy, 1/*b*, is empirically close to the fractal dimension values, $D_q$.

## 4. Questions and discussion

In theory, the fractal parameters are based on entropy functions. The capacity dimension is based on Boltzmann's macro state entropy, the information dimension is based on Shannon's information entropy, and the generalized dimension is based on the second order Renyi's entropy. Along with urban growth, all the entropy values increase, and accordingly, fractal dimension values ascend (Tables 2 and 3). This suggests that both entropy and fractal dimension can be used to describe space filing pattern in the process of city development. Despite the association and similarity, there is significant distinction between spatial entropy and fractal dimension. In urban studies, the entropy values depend on the scale of measurement, and thus we need a set of numbers to characterize a state of urban form. The fractal parameters are based on the concept of scaling, we can use a fractal dimension value to substitute for a number of entropy values. In this sense, fractal theory can provide simpler approach to spatial analysis of cities. According to the empirical analyses, for a given linear scale $\varepsilon$, the numerical relationship between entropy and fractal dimension can be expressed as a linear function such as

$$D_q = a + bM_q, \qquad (9)$$

where $a$ and $b$ are two parameters. Where Beijing is concerned, $a \approx 0$, $b \approx 1/9$; where Hangzhou is concerned, $a \approx 0.1970$, $b \approx -0.4456$. In fact, the spatial entropy and fractal dimension of Beijing's urban form were measured with fixed boxes. That is to say, the largest boxes in different years are the same with one another (Chen and Wang, 2013). However, the state entropy and capacity dimension of Hangzhou city were measured by using variable boxes. The size of the largest boxes changed along with city size in each year (Feng and Chen, 2010). This suggests that the methods of spatial measurement impact on the parameters of equation (9), but the linear relation between entropy and fractal dimension is identifiable.

The value of entropy is related to the state number of a system. It can reflect the diversity of elements in the system. In literature, entropy is often employed to indicate complexity of systems (Cramer, 1993; Pincus, 1991). In fact, entropy is a criterion rather than an index for complex systems. For the distributions with characteristic scale (characteristic length can be found), e.g., urban population density, entropy is an effective measurement for diversity and complicated degree; however, for the distributions without characteristic scale (characteristic length cannot be found), e.g., urban land use pattern, a single entropy value is not enough to measure complexity. In other words, if a system satisfies normal distribution (at least its probability distributions comply with central limit theorem), it can be effectively measured by entropy. In contrast, if a system satisfies power-law distribution (its probability distribution violates central limit theorem), it cannot be easily measured with entropy. In this case, the entropy should be replaced by fractal dimension. A comparison can be drawn as follows (Table 8).

**Table 8 A comparison between entropy meaning and fractal dimension meaning**

| Type | Entropy | Fractal dimension |
| --- | --- | --- |
| **Distribution** | Gaussian (normal) distribution | Pareto-Mandelbrot (power-law) distribution |
| **Scale** | Based on characteristic scale | Based on scaling (scale-free) |
| **System** | Determinate systems | Complex systems |
| **Symmetry** | Spatio-temporal translational symmetry | Scaling symmetry (invariance under contraction or dilation) |
| **State** | Order or chaos | Edge of chaos |
| **Sphere of application** | The patterns and processes with characteristic scale (determinate length, size, mean, eigen value ) | The patterns and processes without characteristic scale (indeterminate length, size, mean, eigen value ) |
| **Example** | Urban population density (Table 6) | Urban land use pattern (Tables 4 and 5) |

The entropy and fractal dimension can be used to measure the spatial homogeneity of urban form, but they can be transformed into the indexes reflecting the spatial heterogeneity. In order to lessen the influence of the linear size of boxes, the index of *entropy ratio* is necessary. Entropy ratio is defined by the ratio of actual entropy to the maximum entropy, that is

$$J = \frac{H}{H_{\max}}, \qquad (10)$$

where $J$ denotes the entropy ratio, $H$ refers to the actual entropy, and $H_{\max}(\varepsilon)$ to the maximum entropy (see the example shown in Table 6). Entropy ratio can also be termed *information quotient* because it is equivalent to the ratio of actual information quantity to the maximum information quantity. Sometimes, the entropy ratio is termed relative entropy (Singh, 2013), but it actually differs from the concept of relative entropy in literature (Cover and Thomas, 1991). Accordingly, the difference between actual entropy and the maximum entropy is termed *information gain* (Batty, 1974; Batty, 1976; Theil, 1967), which can be expressed in the following form

$$I = H_{\max} - H. \qquad (11)$$

Based on entropy ratio or information gain, the degree of *redundancy* can be defined as (Batty, 1974; Reza, 1961)

$$Z = \frac{I}{H_{\max}} = 1 - J = 1 - \frac{H}{H_{\max}}. \qquad (12)$$

Both entropy and entropy ratio (information quotient) reflect the spatial homogeneity, while the redundancy mirrors the spatial heterogeneity of urban form.

Similarly, we can find a set of indexes based on fractal dimension to match entropy ratio, information gain, and redundancy. The corresponding relationships between spatial entropy and fractal dimension can be tabulated as below (Table 4). Fractal dimension can be normalized by the following formula (Chen, 2012b)

$$D^* = \frac{D - D_{\min}}{D_{\max} - D_{\min}}, \tag{13}$$

in which $D^*$ denotes the normalized fractal dimension, $D$ is the actual fractal dimension, $D_{\max}$ represent the maximum fractal dimension, and $D_{\min}$ refers to the minimum fractal dimension. The minimum fractal dimension of urban form can be theoretically treated as 0, thus we have fractal dimension rate such as

$$J^* = \frac{D}{D_{\max}}, \tag{14}$$

where $J^*$ can also be termed fractal dimension quotient or similarity index. Its value comes between 0 and 1, and reflects the spatial homogeneity. Accordingly, an index of dissimilarity can be defined as

$$Z^* = 1 - J^* = 1 - \frac{D}{D_{\max}}, \tag{15}$$

which reflects the spatial heterogeneity of city development. This equation contains an index of fractal dimension difference as below

$$I^* = D_{\max} - D, \tag{14}$$

which reflects the room of urban growth and the scaling exponent of urban density.

The shortcoming of this study lies in the short length of sample path. The number of data points of the two Chinese cities, Beijing and Hangzhou, is only 4 or 5. Fortunately, this is not a critical defect. The short sample path leads to the variability instead of bias. In fact, the well-known Moore's law, which asserts that "the number of transistors in a dense integrated circuit doubles approximately every two years" (from *Wikipedia*, the free encyclopedia), was put forward by means of 5 observational data points and 4 ratios (Moore, 1998). Subsequently this law is consolidated by a greater number of large datasets (Arbesman, 2012; Moore, 2006). A scientific judgment should be given by confidence statement, which comprises *level of confidence* and *margin of error* (Moore, 2009). The confidence level depends on *degree of freedom* rather than sample size. The smaller the sample size, the lower the degree of freedom, and the higher the statistical criterion. Where the above-shown cases are concerned, the squared correlation coefficients suggest approximately perfect fit. The level of confidence is satisfying.

## 5. Conclusions

The significance of fractal dimension can be puzzled out by resolving the problems of entropy, and both entropy and fractal dimension can be used to make spatial analysis of cities. The two measurements can be associated with one another, but there is significant difference. Clarifying the similarities and differences between them is helpful to the appropriate application of entropy and fractal theories in urban studies. The main conclusions of this paper can be reached as follows. **First, the basic approach of measurement of spatial entropy and fractal dimension is the box-counting method.** Using this method, we can find linear relationship between the two measurements. There is subtle difference between the results from the fixed box method and those from variable box method. The similarity of spatial entropy to fractal dimension rests with the following aspects: the two are the measurements of spatial homogeneity, but both of them can be transformed into the measurements of spatial heterogeneity. For the simple fractal structure of cities, the spatial entropy is equivalent in theory to fractal dimension. The fractal dimension seems to be the characteristic value of spatial entropy. In practice, entropy and fractal dimension can be replaced with each other in many cases. For the complex structure of cities, the description based on entropy parameters is complicated, but the fractal dimension description is simple and clear. **Second, the spatial entropy is suitable for describing the urban patterns with characteristic scales.** If we can find effective length, area, volume, average values, eigenvalues, and so on, in a system, the system bears characteristic scales, and spatial entropy is one of the good measurements for the system. These types of systems possess certain Euclidean measures such as perimeter and density, or satisfy the probability distributions with determinative probability structure, including normal distribution, lognormal distribution, Poisson distribution, Gamma distribution, and exponential distribution. For these cases, fractal parameters are inexistent. Typical geographical phenomenon with characteristic scale is urban population density, which follows the exponential decay law. **Third, the fractal dimension is suitable for describing the urban patterns without characteristic scales.** If we cannot find effective length, area, volume, average values, or eigenvalues in a system, the system bears no characteristic scale, or bear scaling property. Fractal dimension is one of the proper scaling exponents for the complex system. These types of systems have no certain Euclidean measures, or satisfy the probability distributions without determinative probability structure. The common scaling distributions include Pareto distribution, Zipf distribution, and Mandelbrot distribution. In short, if a system in a city follows some kind of power-law distributions, it can be described with fractal dimension. For these cases, entropy values are scale-dependent and thus indeterminate. Typical geographical phenomena include urban land use patterns, transport network, and rank-size distribution of cities, which follow the well-known inverse power law of nature.

## Acknowledgements

This research was sponsored by the National Natural Science Foundation of China (Grant No. 41590843 & 41671167). The supports are gratefully acknowledged.

## References

Anastassiadis A (1986). New derivations of the rank-size rule using entropy-maximising methods. *Environment and Planning B: Planning and Design*, 13: 319-334

Arbesman S (2012). *The Half-Life of Facts: Why Everything We Know Has An Expiration Date*. New York: Penguin Group

Batty M (1974). Spatial entropy. *Geographical Analysis*, 6(1): 1-31

Batty M (1976). Entropy in spatial aggregation. *Geographical Analysis*, 8(1): 1-21

Batty M (1992). The fractal nature of geography. *Geographical Magazine*, 64(5): 33-36

Batty M (1995). New ways of looking at cities. *Nature*, 377: 574

Batty M (2008). The size, scale, and shape of cities. *Science*, 319: 769-771

Batty M, Longley PA (1994). *Fractal Cities: A Geometry of Form and Function*. London: Academic Press

Benguigui L, Czamanski D, Marinov M, Portugali Y (2000). When and where is a city fractal? *Environment and Planning B: Planning and Design*, 27(4): 507–519

Bussiere R, Snickers F (1970). Derivation of the negative exponential model by an entropy maximizing method. *Environment and Planning A*, 2(3): 295-301

Chen T (1995). *Studies on Fractal Systems of Cities and Towns in the Central Plains of China* (Master's Degree Thesis). Changchun: Department of Geography, Northeast Normal University (in Chinese)

Chen YG (2008). A wave-spectrum analysis of urban population density: entropy, fractal, and spatial localization. *Discrete Dynamics in Nature and Society*, vol. 2008, Article ID 728420, 22 pages

Chen YG (2012a). The rank-size scaling law and entropy-maximizing principle. *Physica A: Statistical Mechanics and its Applications*, 391(3): 767-778

Chen YG (2012b). Fractal dimension evolution and spatial replacement dynamics of urban growth. *Chaos, Solitons & Fractals*, 45 (2): 115–124

Chen YG (2013). Fractal analytical approach of urban form based on spatial correlation function. *Chaos, Solitons & Fractals*, 49(1): 47-60

Chen YG (2015a). The distance-decay function of geographical gravity model: power law or exponential law? *Chaos, Solitons & Fractals*, 77: 174-189

Chen YG (2015b). Simplicty, complexity, and mathematical modeling of geographical distributions. *Progress in Geography*, 34(3): 321-329 [In Chinese]

Chen YG, Liu JS (2002). Derivations of fractal models of city hierarchies using entropy-maximization principle. *Progress in Natural Science*, 2002, 12(3): 208-211

Chen YG, Wang JJ (2013). Multifractal characterization of urban form and growth: the case of Beijing. *Environment and Planning B: Planning and Design*, 40(5):884-904

Clark C (1951). Urban population densities. *Journal of Royal Statistical Society*, 114: 490-496

Cover TM, Thomas JA (1991). *Elements of Information Theory*. New York: Wiley

Cramer F (1993). *Chaos and order: the complex structure of living systems* (translated by D.I. Loewus). New York : New York: VCH Publishers

Cressie NA (1996). Change of support and the modifiable areal unit problem. *Geographical Systems*, 3 (2–3): 159-180

Curry L (1964). The random spatial economy: an exploration in settlement theory. *Annals of the Association of American Geographers*, 54(1): 138-146

Feder J (1988). *Fractals*. New York: Plenum Press

Feng J (2002). Modeling the spatial distribution of urban population density and its evolution in

Hangzhou. *Geographical Research*, 21(5): 635-646 [In Chinese]

Feng J, Chen YG (2010). Spatiotemporal evolution of urban form and land use structure in Hangzhou, China: evidence from fractals. *Environment and Planning B: Planning and Design*, 37(5): 838-856

Frankhauser P (1998). The fractal approach: A new tool for the spatial analysis of urban agglomerations. *Population: An English Selection*, 10(1): 205-240

Gould PR (1972). Pedagogic review: entropy in urban and regional modelling. *Annals of the Association of American Geographers*, 62(1): 689-700

Grassberger P (1983). Generalized dimensions of strange attractors. *Physics Letters A*, 97(6): 227-230

Grassberger P (1985). Generalizations of the Hausdorff dimension of fractal measures. *Physics Letters A*, 107(1): 101-105

Jiang B, Brandt SA (2016). A Fractal perspective on scale in geography. *International Journal o f Geo-Information*, 5, 95 (doi:10.3390/ijgi5060095)

Jullien R, Botet R (1987). *Aggregation and Fractal Aggregates*. Singapore: World Scientific Publishing Co.

Kwan MP (2012). The uncertain geographic context problem. *Annals of the Association of American Geographers*, 102 (5): 958-968

Longley PA, Batty M, Shepherd J (1991). The size, shape and dimension of urban settlements. *Transactions of the Institute of British Geographers (New Series)*, 16(1): 75-94

Mandelbrot BB (1982). *The Fractal Geometry of Nature*. New York: W. H. Freeman and Company

Mandelbrot BB (1999). *Multifractals and 1/f Noise: Wild Self-Affinity in Physics (1963-1976)*. New York: Springer-Verlag

Moore DS (2009). *Statistics: Concepts and Controversies (7th edition)*. New York: W. H. Freeman and Company

Moore GE (1998). Cramming more components onto integrated circuits. *Proceedings of The IEEE*, 86(1): 82-85 (Reprinted from *Electronics*, pp 114–117, April 19, 1965)

Moore GE (2006). Moore's law at 40. In: D. Brock. *Understanding Moore's Law: Four Decades of Innovation*. Philadelphia: Chemical Heritage Foundation, pp67–84

Openshaw S (1983). *The modifiable areal unit problem*. Norwick: Geo Books

Pincus SM (1991). Approximate entropy as a measure of system complexity. *PNAS*, 88(6): 2297–2301

Rényi A (1961). On measures of information and entropy. *Proceedings of the fourth Berkeley Symposium on Mathematics, Statistics and Probability,1960*, pp547–561

Reza FM (1961). *An Introduction to Information Theory*. NY: McGraw Hill (reprinted in 1994)

Ryabko BYa (1986). Noise-free coding of combinatorial sources, Hausdorff dimension and Kolmogorov complexity. *Problemy Peredachi Informatsii*, 22: 16-26.

Shannon CE (1948). A mathematical theory of communication. *Bell System Technical Journal*, 27(3): 379-423

Singh VP (2013). *Entropy Theory and its Application in Environmental and Water Engineering*. New York: Wiley-Blackwell

Takayasu H (1990). *Fractals in the Physical Sciences*. Manchester: Manchester University Press

Theil H (1967). *Economics and Information Theory*. Amsterdam: North Holland Publishing

Theil H (1967). Economics and Information Theory. Amsterdam: North Holland Publishing

Unwin DJ (1996). GIS, spatial analysis and spatial statistics. *Progress in Human Geography*, 20 (4): 540–551.

Vicsek T (1989). *Fractal Growth Phenomena*. Singapore: World Scientific Publishing Co.

Wheeler JA (1983). Review on *The Fractal Geometry of Nature* by Benoit B. Mandelbrot. *American Journal of Physics*, 51(3): 286-287

White R, Engelen G (1993). Cellular automata and fractal urban form: a cellular modeling approach to the evolution of urban land-use patterns. *Environment and Planning A*, 25(8): 1175-1199

Wilson AG (1968). Modelling and systems analysis in urban planning. *Nature*, 1968, 220: 963-966

Wilson AG (1970). *Entropy in Urban and Regional Modelling*. London: Pion Press

Wilson AG (2000). *Complex Spatial Systems: The Modelling Foundations of Urban and Regional Analysis*. Singapore: Pearson Education